\documentstyle[10pt,draft,psfig]{nature}

\def\kms{km~s$^{-1}$}
\def\la{\ifmmode\stackrel{<}{_{\sim}}\else$\stackrel{<}{_{\sim}}$\fi}
\def\ga{\ifmmode\stackrel{>}{_{\sim}}\else$\stackrel{>}{_{\sim}}$\fi}
\def\fa{\hbox{$.\!\!^{\prime\prime}$}}
\def\fs{\hbox{$.\!\!^{\rm s}$}}

\def\sgra{\mbox{SGR~1806--20}}
\def\sgrb{\mbox{SGR~1900+14}}

\def\src{VLA~J180839--202439}

\newcommand{\titl}{
An expanding radio nebula produced by a giant flare from the magnetar SGR~1806--20
}

\title{\LARGE \bf \titl}
\author{B. M. Gaensler,$^1$
C.~Kouveliotou,$^2$
J.~D.~Gelfand,$^1$
G.~B.~Taylor,$^{3,4}$
D.~Eichler,$^5$
R.~A.~M.~J.~Wijers,$^6$
J.~Granot,$^3$
E.~Ramirez-Ruiz,$^7$
Y.~E.~Lyubarsky,$^5$
R.~W.~Hunstead,$^8$
D.~Campbell-Wilson,$^8$
A.~J.~van der Horst,$^6$
M.~A.~McLaughlin,$^9$
R.~P.~Fender,$^{10}$
M.~A.~Garrett,$^{11}$
K.~J.~Newton-McGee,$^{8,12}$
D. M. Palmer,$^{13}$
N.~Gehrels$^{14}$
and P.~M.~Woods$^{15}$
\\
\scriptsize
$^1$ Harvard-Smithsonian Center for Astrophysics, 60 Garden Street,
Cambridge, MA 02138, USA  \vspace{-2mm}  \\
$^2$ NASA / Marshall Space Flight Center, NSSTC, XD-12, 320 Sparkman
Drive, Huntsville, AL 35805, USA \vspace{-2mm}  \\ 
$^3$ Kavli Institute for Particle Astrophysics and
Cosmology, Stanford University, P.O. Box 20450, Stanford, CA
94309, USA \vspace{-2mm} \\
$^4$ National Radio Astronomy
Observatory, P.O. Box O, Socorro, NM 87801, USA \vspace{-2mm} \\
$^5$ Department of Physics, Ben Gurion University, POB 653,
Beer Sheva 84105, Israel \vspace{-2mm}  \\
$^6$ Astronomical Institute ``Anton Pannekoek'', University
of Amsterdam, Kruislaan 403, 1098 SJ, Amsterdam, The Netherlands \vspace{-2mm} \\
$^7$ Institute for Advanced Study, Einstein Drive,
Princeton, NJ 08540, USA \vspace{-2mm} \\
$^8$ School of Physics, University of Sydney, NSW 2006, Australia  \vspace{-2mm}  \\
$^9$ University of Manchester, Jodrell Bank Observatory, Macclesfield, Cheshire SK11 9DL, UK \vspace{-2mm} \\
$^{10}$ School of Physics and Astronomy, University of Southampton,
Highfield, Southampton SO17 1BJ, UK \vspace{-2mm}  \\
$^{11}$ Joint Institute for VLBI in Europe, Postbus 2, 7990 AA Dwingeloo, The Netherlands \vspace{-2mm}  \\
$^{12}$ Australia Telescope National Facility, CSIRO, PO Box 76,
Epping, NSW 1710, Australia \vspace{-2mm} \\
$^{13}$ Los Alamos National Laboratory, P.O. Box 1663, Los Alamos, NM 87545, USA 
\vspace{-2mm} \\
$^{14}$ NASA / Goddard Space Flight Center, Code 661, Greenbelt, MD 20771, USA \vspace{-2mm} \\
$^{15}$ Universities Space Research Association,
NSSTC, XD-12, 320 Sparkman Drive, Huntsville, AL 35805, USA \vspace{-5mm}}

\headertitle{An Expanding Radio Nebula from \sgra}
\mainauthor{Gaensler et al.}

\begin{document}

\normalsize

\summary{Soft gamma repeaters (SGRs) are ``magnetars'', a
small class of slowly spinning neutron stars with extreme surface
magnetic fields, $B\sim10^{15}$~gauss.\cite{dt92a,kds+98,wt04} On
2004 December 27,  a giant flare\cite{bgm+04} was detected from the
magnetar \sgra,\cite{kds+98} the third such event ever
recorded.\cite{mgi+79,hcm+99} This burst of energy was detected by
a variety of instruments\cite{pbg+05,hbs+05} and even caused an ionospheric
disturbance in the Earth's upper atmosphere recorded around the
globe.\cite{chw+05} Here we report the detection of a fading radio
afterglow produced by this outburst, with a luminosity 500 times
larger than the only other detection of a similar source.\cite{fkb99}
From day~6 to day~19 after the flare from \sgra, a resolved,
linearly polarized, radio nebula was seen, expanding at approximately a
quarter the speed of light.  To create this nebula, at least
$4\times10^{43}$~ergs of energy must have been emitted by the giant
flare in the form of magnetic fields and relativistic particles.
The combination of spatially resolved structure and rapid time
evolution allows a study in unprecedented detail 
of a nearby analog to supernovae and gamma-ray bursts.}

\dates{18 January, 2005}{17 February, 2005}

\maketitle

\begin{figure}
\centerline{\psfig{file=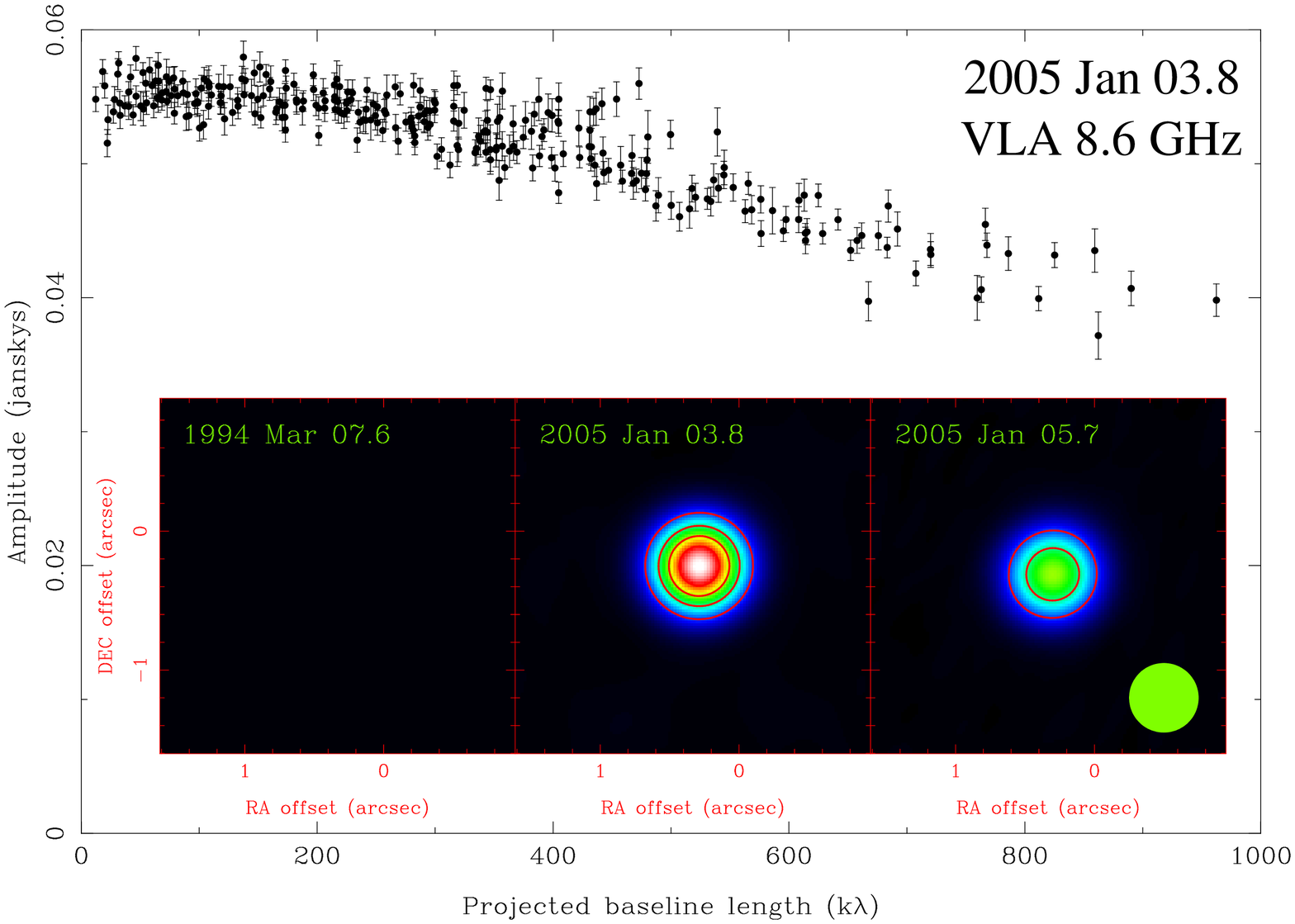,width=\textwidth}}
\caption{}
\label{fig_ext}
\end{figure}

Almost seven days after the 2004~Dec.~27 giant flare, we observed
\sgra\ with the Very Large Array (VLA) in its highest resolution
configuration (maximum baseline length 36.4~km). We identified a
bright but fading radio source designated \src\ (see Fig.~\ref{fig_ext}),
whose position was consistent with the previously reported
localization\cite{kfk+02} of the SGR.  This close juxtaposition,
plus the transient nature of the emission, makes it certain that
\src\ is the radio afterglow of the giant flare from \sgra.  For a
distance\cite{ce04} to \sgra\ of $15d_{15}$~kiloparsecs, the  1.4-GHz
flux density of this source at first detection implies an isotropic
spectral luminosity of $5d_{15}^2\times10^{15}$~W~Hz$^{-1}$,
approximately 500 times larger than the radio afterglow seen from
\sgrb\ after a giant flare in 1998.\cite{fkb99} No other magnetar
has been detected in the radio band, either in quiescence or during
active periods.\cite{lx00,ktw+01}

Given the very bright nature of this afterglow, we 
organized an international campaign over a broad range of frequencies,
0.35 to 16~GHz, to track the decay of the radio emission of \src. Here
we present a subset of these observations, made on days 6 to 19 after the
giant flare, consisting of images made using the VLA,
the Australia Telescope Compact Array (ATCA), the Westerbork Synthesis
Radio Telescope (WSRT) and the Molonglo Observatory Synthesis Telescope
(MOST) (see Supplementary Methods for further information).  

\begin{figure}[h!]
\centerline{\psfig{file=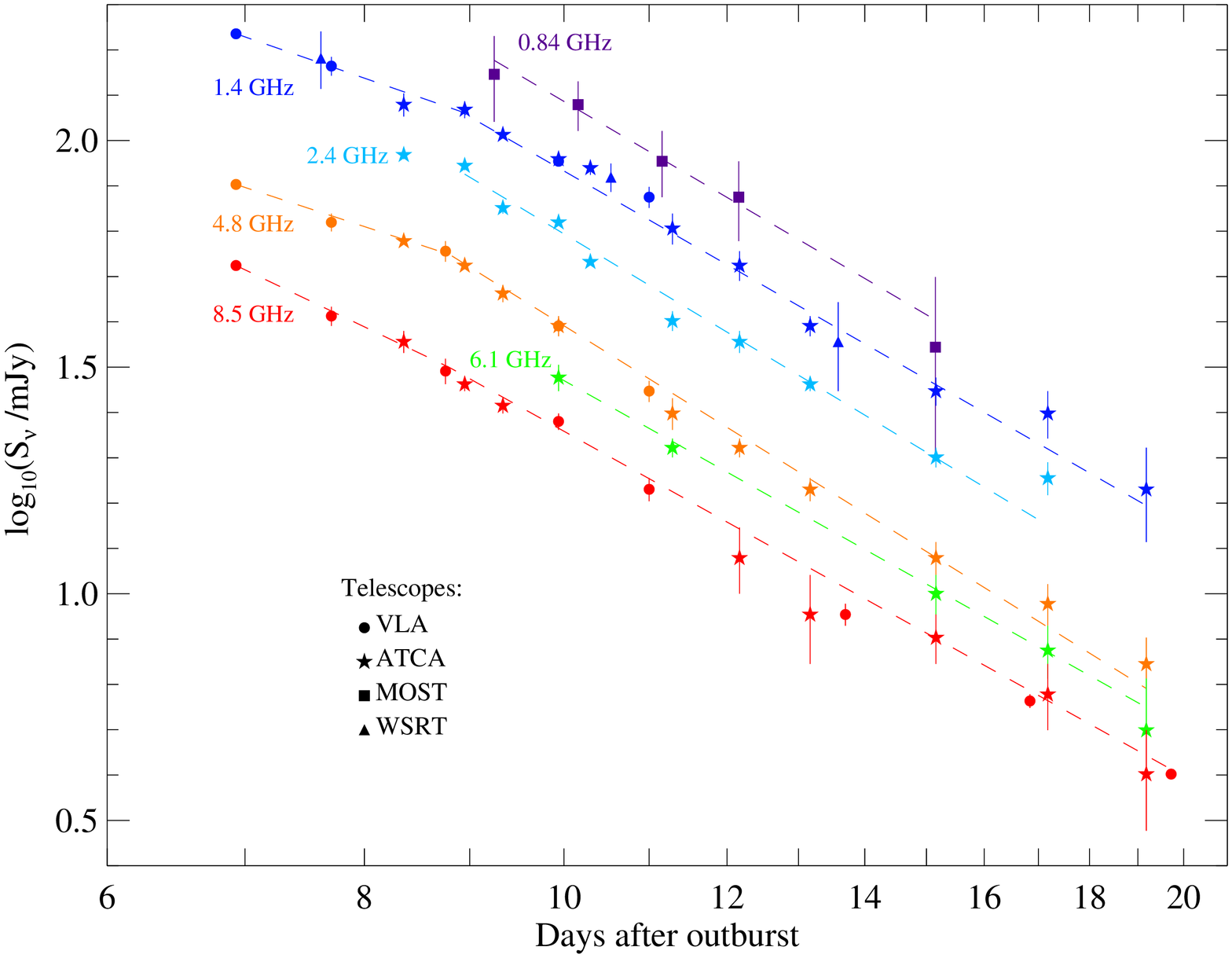,width=\textwidth,angle=90}}
\caption{}
\label{fig_lc}
\end{figure}

Figure~\ref{fig_lc} shows the combined light curves from these four
telescopes covering the frequency range 0.84 to 8.5~GHz.  These
data are consistent with a sudden increase in the decay rate at day
8.8, as summarised in Table~\ref{tab_lc} and shown by the linear
fits in Figure~\ref{fig_lc}. Specifically, if we assume that $S_\nu
\propto t^{\delta}$ (where $S_\nu$ is the flux density at
frequency $\nu$), after day~8.8 we find an achromatic and rapid
decline, $\delta \approx -2.7$, in  six independent frequency bands
(a similarly rapid decline was also observed\cite{fkb99} for the
radio afterglow of \sgrb\ in 1998).  
After carefully accounting for the instrumental response of the VLA
antennas we find that \src\ is significantly linearly polarized
(see Fig.~\ref{fig_vstime}), which indicates that the emission
mechanism is synchrotron radiation.  In our earliest observations this emission
was already optically thin, but showed clear evidence for a spectral
steepening at high frequencies (see caption to Fig.~\ref{fig_lc}).
From day~11.2 onward, the spectrum was consistent
with an unbroken power law from 0.84 to 8.5~GHz with $\alpha =
-0.75\pm0.02$ (where $S_\nu \propto \nu^\alpha$), again similar to
the 1998 afterglow of \sgrb.  This implies a power-law energy
distribution of the emitting electrons, $dN/dE\propto E^{-p}$, with $p
= 1- 2\alpha  = 2.50 \pm 0.04$.

Our highest resolution measurements are those made with the VLA at
8.5~GHz.  The visibility data from these observations, as shown for
one epoch in Figure~\ref{fig_ext}, demonstrate that \src\ is resolved.
A Gaussian is a good fit to the visibilities at each epoch, with
no significant persistent residuals (forthcoming higher resolution
images from the VLBA and MERLIN will test the validity of this model).
Figure~\ref{fig_vstime} demonstrates that from day~6.9 to day~19.7, the
data were consistent with constant isotropic expansion since outburst at
a speed $v/c = (0.27\pm 0.10)d_{15}$, with no noticeable deceleration
as of day~19.7.  Other than in one observation at day 16.8, the source
was significantly elliptical, with an axial ratio $\sim0.6$ and with
the major axis oriented $\approx 60^\circ$ west of north.

The spectrum and angular size of \src\ allow us to apply standard
equipartition arguments for synchrotron sources,\cite{pac70}
implying a minimum magnetic field $B_{\rm min} = 0.02
d_{15}^{-2/7}[(1+\kappa)F_{100}/f]^{2/7}\theta_{50}^{-6/7}$~gauss,
where $100F_{100}$~mJy is the flux density of \src\ at 1.4~GHz,
$\kappa$ the ratio of the energy in heavy particles to that in
electrons, $f$ is the volume filling factor of magnetic fields and
relativistic particles, and $50\theta_{50}$~mas is the source's
angular diameter. The minimum energy in particles and magnetic
fields in the emitting region is $E_{\rm min}=4\times10^{43}~
d_{15}^{17/7}[(1+\kappa)F_{100}]^{4/7}f^{3/7}\theta_{50}^{9/7}$~ergs.
The spectra show no evidence of self-absorption at frequencies above
0.6~GHz at early times,\cite{cha05} which is consistent with these
parameters provided that the emitting medium has a density $n_0 \la
0.1f$~cm$^{-3}$.  We can derive an additional independent energy estimate
because of this rare opportunity to measure the expansion velocity
directly.  From the constant expansion speed observed over the first 20
days, we infer $ E_{\rm min} \approx 6 \times 10^{42}~ \Omega
(n_0/0.01~{\rm cm}^{-3})
(v/0.27c)^5 $~ergs, where $\Omega$ is the opening  solid
angle of the ejected material.

Giant flares from magnetars are thought to result from shearing and
reconnection of the extreme magnetic fields near the neutron star
surface.\cite{td01,td95} The inferred minimum energy in the radio nebula is
somewhat smaller than the emitted gamma-ray energy,\cite{pbg+05,hbs+05}
but is much larger than the electron/positron pair luminosity that would be
expected to survive annihilation close to the magnetar. This suggests that
baryons may have been ablated off the surface by the intense illumination
of the flaring magnetosphere.\cite{td01,td95} The radio nebula could be
naturally created by these baryons, which move off the magnetar at high
velocity, $\ga0.5c$, and then shock the ambient medium.

The very steep decay of the radio emission after day~8.8, $\delta \approx
-2.7$, combined with the observed sub-luminal expansion velocity of
\src, is difficult to produce in standard gamma-ray burst blast-wave
models.\cite{rho99,gps99,cw03} The light curves may thus represent an
adiabatically expanding population  of electrons accelerated at a particularly active
phase, such as might occur if the ejecta collided with a pre-existing
shell.  Such a shell is naturally made by \sgra\
itself, since its quiescent wind\cite{tdw+00} of luminosity\cite{wkg+02}
$\sim 10^{34}$~ergs~s$^{-1}$ will sweep up a bow shock\cite{wil96,gjs02}
of stand-off distance $\sim10^{16}$~cm (corresponding to an angular
extent $\sim40d_{15}^{-1}$ mas) as it moves through the interstellar
medium (ISM) at a typical neutron star velocity\cite{acc02} of
$\sim200$~\kms. The star's motion creates a cigar-shaped cavity, mostly
as a wake that trails the bow shock.  If this pre-existing shell is hit
by $\sim10^{43}$--$10^{44}$~ergs of energy from the SGR's giant flare,
it will be shocked and swept outward, resulting in 
a violent episode of particle acceleration that puts much of the energy into
a steadily expanding synchrotron-emitting shell 4 to 8 days after the giant
flare.  If we suppose that this shell maintains constant thickness
and constant expansion speed, then its volume,
$V$, increases as $t^2$, and the magnetic field will decay as $V^{-1/2}$
if directed within the tangent plane of the shell
($V^{-2/3}$ if tangled in three dimensions). This predicts
a power law index for the radio decay $\delta =
(7\alpha-3)/3 = -2.75\pm0.05$ (for $B \propto V^{-1/2}$) or $\delta =
(8\alpha -4)/ 3 = -3.33\pm0.05$ (for $B \propto V^{-2/3}$), consistent
with the steep decay observed here.
The overall evolution can be complicated by the fact that the ejecta may
be somewhat collimated, and may hit the shell at different places and
times --- we defer detailed modelling to later papers, pending higher
resolution images from MERLIN and the VLBA.

At early times, the polarization position angle on the sky is
approximately perpendicular to the axis of the radio source (see
Fig.~\ref{fig_vstime}), suggesting that the magnetic field in the
emitting plasma (on average) is aligned preferentially along this
axis. This is consistent with the shock producing
a preferred magnetic anisotropy in the shock plane.
Between observations at days 11.0 and 13.7  the
polarized fraction and polarization angle both changed noticeably
and a possible bump in the 1.4-GHz light curve is apparent; at day
16.8, the position angle of the major axis of the source also changed
considerably.  These results suggest that a different part of the
outflow may have assumed the dominant role in the emission, as can
occur if one region fades faster than another.

The intensity of this radio afterglow confirms the conservative inference
made from the  X- and gamma-ray detections\cite{pbg+05,hbs+05} that this
event was $\ga2$ orders of magnitude more energetic than the 1998
giant flare from \sgrb. It is difficult to attribute the difference to
beaming effects, because the measured expansion velocity ($\sim 0.3c$)
appears to be modest.  A release of $>10^{46}$~ergs in a single
magnetar flare\cite{pbg+05,hbs+05} suggests that a rather large fraction, $\sim
10$ percent, of the total magnetic  energy can be released at once.
Continued measurements of the morphology of the expanding radio source
can provide an indication of whether the energy release took place at
a specific location on the star's surface, or was  a truly global phenomenon
that rearranged the crust or even the entire interior.

\bibliographystyle{nature}
\bibliography{journals,modrefs,psrrefs}

\bigskip
\noindent
Supplementary Information accompanies the paper on www.nature.com/nature.

\bigskip
\noindent
Correspondence and requests for materials should be
addressed to B.M.G.  (bgaensler@cfa.harvard.edu).

\begin{acknowledge}

\small

We thank Jim Ulvestad, Joan Wrobel, Bob Sault, Tony Foley and Rene
Vermeulen for rapid scheduling of the VLA, ATCA and WSRT; Tracey
DeLaney, Ger de Bruyn  and Crystal Brogan for assistance with data
analysis; and Dick Manchester, Dale Frail and Mark Wieringa for help
with the observations.  NRAO is a facility of the NSF operated under
cooperative agreement by AUI.  The Australia Telescope is funded by
the Commonwealth of Australia for operation as a National Facility
managed by CSIRO.  The MOST is operated by the University of Sydney and
supported in part by grants from the ARC.  The WSRT is operated by ASTRON
with financial support from NWO.  B.M.G. acknowledges the support of
NASA through a Long Term Space Astrophysics grant.  D.E. acknowledge
support from the Israel-U.S. BSF, the ISF, and the Arnow Chair of
Physics. Y.L. acknowledges support from the German-Israeli Foundation.
R.A.M.J.W and A.J.H. acknowledge support from NWO.

\end{acknowledge}

\normalsize

\clearpage

\centerline{   }

\vspace{0.5cm}
\centerline{\bf FIGURE \& TABLE CAPTIONS}

\vspace{0.5cm}

\noindent {\bf TABLE~\ref{tab_lc}:} The rate of decay of the radio
emission from \src\ at six independent frequencies. At each frequency,
$\nu$, it has been assumed that the radio flux density decays as $S_\nu
\propto t^{\delta_\nu}$, with a break in the power law index,
$\delta_\nu$, at time $t_0$. To determine values of $t_0$ and
$\delta_\nu$, a weighted least squares fit of a broken power law
has been applied to each data-set, with $t_0$ a free parameter. In
each case, the fit shown is the only local minimum in $\chi^2$ which
meets the requirements that there are at least two data-points on
either side of the break, the change in temporal index on either
side of the break is larger than its uncertainties, and the power-law
fits on either side of the break meet at the break point.  Before
day 8.8, we find that $\delta_\nu$ possibly decreases with $\nu$;
after day 8.8, the flux decays rapidly at all frequencies with
a power-law index $\delta \approx -2.7$, independent of $\nu$.

\vspace{0.5cm}

\noindent {\bf FIGURE~\ref{fig_ext}:} Radio emission from \src\ at
8.5~GHz.  The main panel shows the visibility amplitude as a function
of projected baseline length (in units of thousands of wavelengths;
100~k$\lambda \approx 3.5$~kilometers) at epoch 2005~Jan~03.8 (6.9
days after the giant flare), as seen by the VLA.  The data have been
self-calibrated in phase until the solution converged, and each baseline
has then been time-averaged over the entire observation of duration
40~minutes.  The error bars show the standard error in the mean of the
amplitude on each baseline.  The decrease in amplitude as a function of
increasing baseline length clearly indicates that the source is resolved.
The inset shows the image of the source at three epochs, smoothed to a
uniform resolution of $0\fa5$ (indicated by the green circle at lower
right).  The origin of the coordinate axes is the position of \sgra\
measured with the {\em Chandra X-ray Observatory},\cite{kfk+02} which
has an uncertainty of $0\fa3$ in each coordinate.  The false-colour
representation is on a linear scale, ranging from --0.3 to the peak
brightness of 53~mJy~beam$^{-1}$. The contours are drawn
at levels of 20\%, 40\%, 60\% and 80\% of this peak.  No source is
seen in archival 8.5-GHz data from 1994~March, down to a 5-$\sigma$
upper limit of 0.1~mJy.  In the days after the giant flare,
a bright but rapidly fading source is now seen at this position.
The precise location of \src\ was determined by phase referencing to
several nearby calibrators with well-determined positions.  Our best
measurement was on Jan.~16.6, for which we measured a position for
\src\ (equinox J2000) of Right Ascension (R.A.) $18^{\rm h}08^{\rm
m}39\fs343\pm0\fs002$, Declination (Dec.) $-20^\circ24'39\fa80\pm0\fa04$.
The source's proper motion over the time span presented in this
paper is $-2.8\pm6.5$~mas~day$^{-1}$ in R.A. and
$-2.2\pm6.5$~mas~day$^{-1}$ in Dec.

\vspace{0.5cm}

\noindent {\bf FIGURE~\ref{fig_lc}:} Time evolution of the radio
flux density  from \src.  The x-axis indicates days since the giant
flare was detected from \sgra, on 2004~Dec~27.90 UT.  The radio
data originate from ATCA, MOST and WSRT measurements made in six
independent frequency bands.  Each symbol represents a different
telescope, while each colour indicates a different frequency.
Measurement uncertainties are indicated at the 1-$\sigma$ level.
Fits to the data are indicated by dashed lines, and represent the
results of applying the broken power-law model described in
Table~\ref{tab_lc} to the data.  Significant deviations from this
fit are seen at both 1.4 and 2.4~GHz, suggesting short-term
time-variability in the source (most notably the possible ``bump''
in the 1.4~GHz light curve seen with multiple telescopes on days
10--11 after the flare).
These data also allow us to compute the evolution of the radio
spectral index, $\alpha$ (defined as $S_\nu \propto \nu^\alpha$).
At three epochs with good frequency coverage between 8.4 and 9.9
days after the flare, there is clear evidence for a spectral break,
from $\alpha \approx -0.66$ below $\sim5$~GHz to $\alpha \approx
-1.0$ above. Other data cannot rule out this break being present
from day 6.9 (when the source was first detected) through to day~11.0.
From day~11.2 onward, the spectrum has been consistent with an
unbroken power law from 0.84 to 8.5~GHz with $\alpha = -0.75\pm0.02$.

In addition to the data shown here, on 2004~December~29, we used
the Parkes Radio Telescope at 1.4~GHz to search for radio pulsations
from \sgra.  For dispersion measures in the range 0 to
2000~parsecs~cm$^{-3}$, we found no pulsed signal at or near the
star's X-ray period\cite{kds+98} of 7.5~sec down to a level of
$\approx0.2$~mJy (these data provide no constraint on the unpulsed
flux).

\vspace{0.5cm}

\noindent {\bf FIGURE~\ref{fig_vstime}:} Structural and polarization
properties of \src\ as a function of time, as seen with the VLA at
8.5~GHz.  The x-axis indicates days since the giant flare.  The
uppermost panel plots the radius of the source, determined by
modelling\cite{smk+93} the visibilities at each epoch as a
two-dimensional Gaussian function in the Fourier plane of arbitrary
position, amplitude, diameter, axial ratio and orientation, and
then taking the geometric mean of the semi-major and semi-minor
axes.  The broken line shows a weighted linear least squares fit
to the data.  The indicated expansion velocity assumes two-sided
or isotropic expansion at a distance of 15~kiloparsecs.  The two
panels below this show the axial ratio and position angle (measured
north through east) of this best-fit Gaussian at each epoch.  The
fourth panel shows the fractional linear polarization of \src\ at
8.5~GHz. We find that the position angle of this linearly polarized
emission is a linear function of $\nu^{-2}$ at each epoch, indicating
the presence of Faraday rotation from foreground magnetised plasma.
We measure a Faraday rotation measure (RM) of $+272\pm10$~rad~m$^{-2}$
at multiple epochs, similar to the value RM~$=+290\pm20$~rad~m$^{-2}$
obtained for the adjacent calibrator, MRC~B1817--254, and typical
of RMs seen through the Galactic plane.\cite{btj03}  Any contribution
to the RM from the immediate environment of the magnetar must thus
be small.  The fifth panel shows the position angle of the electric
field vector of linear polarization from \src\, after correction
for this foreground Faraday rotation.  Uncertainties at the 1-$\sigma$
level are indicated for all data.

\clearpage

\centerline{   }

\vspace{0.5cm}

\begin{table}[h!]
\caption{ }
\label{tab_lc}
\begin{tabular}{cccc} \hline
$\nu$ (GHz)   & $t_0$ (days) & $\delta_\nu$ ($t < t_0$) & $\delta_\nu$ ($t >t_0$) \\
\hline
0.84  & $\le10.2$ & $\ldots$ & $-2.7\pm0.8$ \\
1.4  & $9.0^{+0.4}_{-0.6}$ & $-1.6\pm0.2$ & $-2.61\pm0.09$ \\
2.4  &  $\le9.0$ & $\ldots$  & $-2.74\pm0.07$ \\
4.8  & $8.8^{+0.2}_{-0.4}$ & $-1.5\pm0.1$ & $-2.84\pm0.08$ \\
6.1  & $\le11.3$ & $\ldots$ & $-2.6\pm0.2$ \\
8.5  & $8.8^{+0.2}_{-0.4}$ & $-2.2\pm0.2$ & $-2.54\pm0.04$ \\ \hline
\end{tabular}
\end{table}

\vspace{0.5cm}

\clearpage

\begin{figure}
\centerline{\psfig{file=fig_vstime.eps,width=\textwidth,angle=270}}
\caption{}
\label{fig_vstime}
\end{figure}

\clearpage

%\centerline{   }

\vspace{-1cm}
\centerline{\bf SUPPLEMENTARY METHODS}
%\vspace{-1cm}

\setcounter{table}{0}

\begin{table}[hb]
\footnotesize
\vspace{-5mm}
\caption{\footnotesize Flux density of \src, the transient radio source
coincident with \sgra, as a function of both time and frequency, as
measured by the VLA, ATCA, WSRT and MOST.
The epoch of the flare was
2004~Dec~27.89 UT.  Uncertainties in each flux measurement are given
at the 1-$\sigma $ level.  All telescopes carried out interferometric
synthesis imaging of the field of \sgra\ to make these measurements. The
absolute flux scales were set using bright standard calibrators;
on several days, different telescopes observed the source near
simultaneously and obtained near identical flux  measurements, indicating
the reliability of this calibration. Phase calibration was carried using
regular observations of nearby bright compact sources. At times when \src\
was bright enough, self-calibration in phase only was also applied. The
data were then Fourier transformed to the image plane, deconvolved using
the point response derived from the synthesis transfer function, and then
convolved with a Gaussian beam corresponding to the diffraction limited
resolution. Fluxes were then extracted in three ways: by integrating
the surface brightness of the source, by fitting the image of the source
to a Gaussian, and by modelling the source as a Gaussian in the complex
visibility plane --- results for total fluxes were consistent among these
three approaches.  The data with lower spatial resolution (most notably
the MOST and WSRT observations) suffered from confusion from the
bright radio source VLA~J180840--202441 [Vasisht, Frail \& Kulkarni, 1995,
{\em The Astrophysical Journal (Letters)}, {\bf 440}, L65; Frail, Vasisht
\& Kulkarni, 1997, {\em The Astrophysical Journal (Letters)}, {\bf 480},
L129], $14''$ to the east of \sgra, associated with the Luminous Blue
Variable (LBV) 1806--20 [Hurley et al., 1999, {\em The Astrophysical
Journal (Letters)}, {\bf 523}, L37].  For these data, difference imaging
and background subtraction were carefully applied to extract the radio
flux of the transient source.  The estimated uncertainties account for
the systematic effects associated with this approach.}
\vspace{1mm}
\begin{center}
\begin{tabular}{ccccccccc} \hline
Mean Epoch & Days after & Telescope & 
\multicolumn{6}{c}{Flux Density (mJy)} \\
(UT) & {Flare} & {} & {0.84} & {1.4} & {2.4} & 
{4.8} & {6.1} & {8.5} \\
{} & {} & {} & {GHz} & {GHz} & {GHz} & {GHz} & {GHz} & {GHz} \\
\hline
Jan~03.83 & 6.93 & VLA & \ldots & 172$\pm$4 & \ldots & 80$\pm$1 & 
\ldots & 53$\pm$1 \\
Jan~04.52 & 7.62 & WSRT &\ldots & 152$\pm$22 & \ldots & \ldots & \ldots
 & \ldots \\
Jan~04.61 & 7.71 & VLA & \ldots & 146$\pm$7 & \ldots & 66$\pm$3 & 
\ldots & 41$\pm$2 \\
Jan~05.26 & 8.36 & ATCA & \ldots & 120$\pm$7 & 93$\pm$2 & 60$\pm$1 & 
\ldots & 36$\pm$2 \\
Jan~05.66 & 8.76 & VLA & \ldots & \ldots & \ldots & 57$\pm$3 & \ldots
 & 31$\pm$2 \\
Jan~05.85 & 8.95 & ATCA & \ldots & 117$\pm$5 & 88$\pm$2 & 53$\pm$1 & 
\ldots & 29$\pm$1 \\
Jan~06.15 & 9.25 & MOST & 140$\pm$30 & \ldots & \ldots & \ldots & \ldots
 & \ldots \\
Jan~06.24 & 9.34 & ATCA & \ldots & 103$\pm$2 & 71$\pm$2 & 46$\pm$2 & 
\ldots & 26$\pm$1 \\
Jan~06.85 & 9.95 & ATCA & \ldots & 91$\pm$2 & 66$\pm$1 & 39$\pm$2 & 
30$\pm$2 & \ldots \\
Jan~06.85 & 9.95 & VLA & \ldots & 90$\pm$2 & \ldots & 39$\pm$1 & \ldots
 & 24$\pm$1 \\
Jan~07.06 & 10.16 & MOST & 120$\pm$15 & \ldots & \ldots & \ldots & 
\ldots & \ldots \\
Jan~07.20 & 10.30 & ATCA & \ldots & 87$\pm$3 & 54$\pm$1 & \ldots & 
\ldots & \ldots \\
Jan~07.44 & 10.54 & WSRT & $\ldots$ & 83$\pm$6 & \ldots & \ldots & 
\ldots & \ldots \\
Jan~07.90 & 11.00 & VLA & \ldots & 75$\pm$4 & \ldots & 28$\pm$2 & 
\ldots & 17$\pm$1 \\
Jan~08.06 & 11.16 & MOST & 90$\pm$15 & \ldots & \ldots & \ldots & \ldots
 & \ldots \\
Jan~08.19 & 11.29 & ATCA & \ldots & 64$\pm$5 & 40$\pm$2 & 25$\pm$2 & 
21$\pm$1 & \ldots \\
Jan~09.06 & 12.16 & MOST & 75$\pm$15 & \ldots & \ldots & \ldots & \ldots
 & \ldots \\
Jan~09.07 & 12.17 & ATCA & \ldots & 53$\pm$4 & 36$\pm$2 & 21$\pm$1 & 
\ldots & 12$\pm$2 \\
Jan~10.07 & 13.17 & ATCA & \ldots & 39$\pm$2 & 29$\pm$1 & 17$\pm$1 & 
\ldots & 9$\pm$2 \\
Jan~10.49 & 13.59 & WSRT & $\ldots$ & 36$\pm$8 & \ldots & \ldots & 
\ldots & \ldots \\
Jan~10.60 & 13.70 & VLA & \ldots & \ldots & \ldots & \ldots & \ldots & 
9.0$\pm$0.5 \\
Jan~12.05 & 15.15 & MOST & 35$\pm$15 & \ldots & \ldots & \ldots & \ldots
 & \ldots \\
Jan~12.06 & 15.16 & ATCA & \ldots & 28$\pm$2 & 20$\pm$1 & 12$\pm$1 & 
10$\pm$1 & 8$\pm$1 \\
Jan~13.74 & 16.84 & VLA & \ldots & \ldots & \ldots & \ldots & \ldots & 
5.8$\pm$0.2 \\
Jan~14.08 & 17.18 & ATCA & \ldots & 25$\pm$3 & 19$\pm$2 & 10$\pm$1 & 
8$\pm$1 & 6$\pm$1 \\
Jan~16.08 & 19.18 & ATCA & \ldots & 17$\pm$4 & \ldots & 7$\pm$1 & 
5$\pm$2 & 4$\pm$1 \\
Jan~16.62 & 19.72 & VLA & \ldots & \ldots & \ldots & \ldots & \ldots & 
4.0$\pm$0.1 \\ \hline
\end{tabular}
\end{center}
\end{table}

\end{document}